# Acceleration of Histogram-Based Contrast Enhancement via Selective Downsampling

Gang Cao[1*], Huawei Tian[2], Lifang Yu[3], Xianglin Huang[1] and Yongbin Wang[1]

[1]School of Computer Science, Communication University of China, Beijing, China

[2]People's Public Security University of China, Beijing, China

[3]School of Information Engineering, Beijing Institute of Graphic Communication, Beijing, China

Correspondence author: gangcao@cuc.edu.cn

## Abstract

In this paper, we propose a general framework to accelerate the universal histogram-based image contrast enhancement (CE) algorithms. Both spatial and gray-level selective down-sampling of digital images are adopted to decrease computational cost, while the visual quality of enhanced images is still preserved and without apparent degradation. Mapping function calibration is novelly proposed to reconstruct the pixel mapping on the gray levels missed by downsampling. As two case studies, accelerations of histogram equalization (HE) and the state-of-the-art global CE algorithm, i.e., spatial mutual information and PageRank (SMIRANK), are presented detailedly. Both quantitative and qualitative assessment results have verified the effectiveness of our proposed CE acceleration framework. In typical tests, computational efficiencies of HE and SMIRANK have been speeded up by about 3.9 and 13.5 times, respectively.

## 1. Introduction

Contrast enhancement (CE) of digital images refers to the operations which improve the perceived contrast. Such contrast is typically defined as the dynamic range of pixel gray-levels within global or local image regions. CE is a widely used image enhancement tool in real applications [1]. Generally, a good CE technique is expected to have: 1) more contrast improvement with less image distortion; 2) low computational cost.

In consideration of its importance in image processing, plenty of previous works have presented image CE techniques. In terms of the mapping applied to pixel gray-levels, the existing CE algorithms can be generally categorized as global, local and hybrid ones [2]. Global CE modifies an image via an identical pixel value mapping, such that the gray-level histogram of the processed image resembles the desired one and becomes more spread than that of the original image [2, 3]. Local CE improves contrast by altering pixels in terms of local properties, and typically operates in the image transform domains, such as the discrete cosine transform (DCT) [4] and the discrete wavelet transform (DWT) [5]. Local CE can also be enforced by adaptively applying the global CE to local image regions [3]. Hybrid CE, which combines the global and local CE together, can improve the unified perception of both global and local contrasts [6].

Note that most of existing global CE techniques need to use the gray-level or transform coefficient histogram of input images. As summarized in [3], histogram modification-based CE received the most attention due to straightforward and intuitive implementation qualities. One popular global CE method is histogram equalization (HE) [1], which improves contrast by redistributing the probability density of gray-levels towards uniformity. The prominent merit of HE lies in its high computational efficiency. However, HE might incur excessive enhancement and unnatural artifacts on the images with high peaks in histograms. In order to attenuate such deficiency, lots of improved HE algorithms [3, 7-11] have been proposed, where the histogram modification framework (HMF) [3] is an influential one. HMF treats CE as an optimization problem by minimizing a cost function which includes the penalty of the histogram deviation from original to uniform. Gu *et al.* [10] proposed an optimal histogram mapping for automatic CE based on a novel reduced reference image quality metric for contrast change. In [11], a complete HMF is presented by integrating the automatic parameter selection via saliency preservation. T. Celik [6] proposed spatial entropy based CE (SECE) by novelly incorporating the spatial distribution characteristics of pixels into the design of gray-level mapping function. SECE can always slightly improve image contrast without incurring serious image quality degradation. Recently, T. Celik [2] proposes the state-of-the-art global image CE method, SMIRANK, by using spatial mutual information of pixels and PageRank. Although good enhancement quality is achieved, such a algorithm runs rather slower than most of other CE algorithms including HE, HMF, SECE and the adaptive gamma correction with weighting distribution (AGCWD) [12]. Besides, the transform coefficient histogram has also been used in CE design [4].

Low computational complexity is an important requirement for the real applications of CE techniques, such as those in low-power embedded imaging systems and the internet of things. Moreover, existing histogram-based CE algorithms may become computationally inefficient in enhancing the images with large spatial resolution and high dynamic range (HDR). In such frequently encountered scenarios, both the one- and two-dimensional gray level histograms involved calculations would become highly time-consuming. Therefore, it is essential and significant to speed up general histogram-based CE algorithms. However, despite some particularly designed fast CE methods [13, 14], to the best of our knowledge, there does not exist any prior work focusing on the acceleration of general histogram-based CE. In this work, we study such a universal problem by proposing a general acceleration framework to improve the existing CE algorithms themselves, instead of applying parallel computing or device-dependent computational strategies [15, 16]. Selective downsampling in both spatial and gray-level domains is employed to decrease computational cost, while the visual quality of enhanced images is still preserved. As case studies, the accelerations of HE and SMIRANK are presented detailedly.

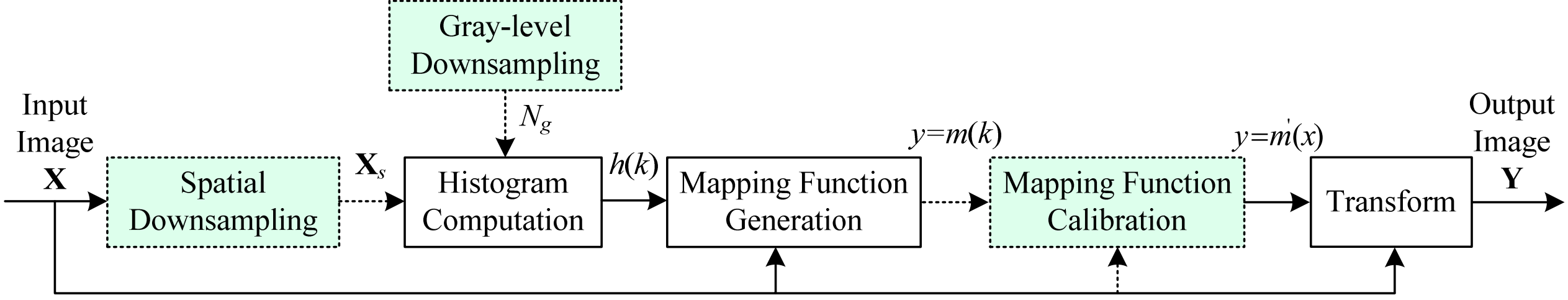

**Fig. 1.** The proposed CE acceleration framework. The new operations integrated into a general histogram-based CE are plotted in dotted-line boxes.

The remainder of this paper is organized as follows. Section 2 proposes the acceleration framework for general histogram-based CE techniques. Section 3 presents the detailed case studies on HE and SMIRANK, followed by the experimental results and discussion given in Section 4. The conclusions are drawn in Section 5.

## 2. Proposed CE Acceleration Framework

As illustrated by the solid-line boxes of Fig.1, a general gray level histogram-based CE technique typically consists of three basic steps: histogram computation, mapping function generation and image transform. Specifically, a single global or dense local histogram(s) of an input image is first computed. Then an elaborately designed mapping function is derived from such statistics and the image. Lastly, the pixel gray-level mapping is globally applied to the input image.

In order to accelerate such histogram-based CE processing, we propose to speed up the histogram computation by selective spatial and gray level downsampling. The yielded low-resolution histogram would benefit the fast generation of a low-resolution mapping function, which is merely defined in the quantized gray levels. However, such a mapping function is improper to be directly applied to neither downsampled nor primary input images, because the abnormal stratification artifacts are incurred and the function on missed gray levels are undefined. As another contribution of this framework, we novelly propose to calibrate such a low-resolution mapping function so that it reasonably covers the full gray level dynamic range of input images.

### *2.1 Fast Histogram Computation via Selective Downsampling*

Histogram construction is an essential and fundamental step in general histogram-based CE methods, which can be accelerated by decreasing the computational costs of histogram computation and histogram-involved operations in mapping function generation. As such, the cost-effective selective spatial and gray-level downsampling is proposed to be used as acceleration strategies, as plotted in the dotted-line boxes of Fig. 1.

In order to decrease the number of counted pixels, spatial downsampling is first applied to the $B$-bit input grayscale image denoted by $\mathbf{X}(i, j)$, $i$=0, 1, …, $M$-1, $j$=0, 1, …, $N$-1, where [$M$, $N$] denote the image size. Without loss of generality, the uniform downsampling with the sampling step $s$ is used to diminish the computational cost. The spatially downsampled image $\mathbf{X}_s$ is generated as

$$\mathbf{X}_s(i, j)=\mathbf{X}(s \cdot i,\ s \cdot j) \tag{1}$$

where $i$=0, 1, …,$\lfloor M/s \rfloor$-1, $j$=0, 1, …, $\lfloor N/s \rfloor$-1, and $\lfloor \cdot \rfloor$ denotes rounding towards zero. If $s$ is limited within a proper range, gray-level histogram of the downsampled image can keep consistent shape as that of the input. The integer, instead of fraction, datatype of $s$ values are adopted to decrease the additional computational cost incurred by downsampling. $s$ can also be larger for large size of images due to the higher correlationship between local pixels.

The gray-level histogram of $\mathbf{X}_s$ can be obtained as

$$h(k) = \sum_{i=0}^{\lfloor M/s \rfloor -1} \sum_{j=0}^{\lfloor N/s \rfloor -1} 1\left( \left\lfloor \frac{\mathbf{X}_s(i, j)}{\Delta} \right\rfloor = k \right) \tag{2}$$

where $k$=0, 1, ..., $N_g$-1. $\Delta$ is the quantization step of gray levels. $N_g = 2^B/\Delta$ is the number of histogram bins. $1(\cdot)$ is an indicator function. $h(k)$ would be used to yield the mapping function, where post operations typically run on such a histogram. As such, the histogram dimension, $N_g$, also affects the computational cost of CE algorithms. Here, we propose to decrease the statistical precision of histograms properly by reducing the number of bins. Such histogram coarsening can be treated as the gray-level downsampling of images.

Overall, we can see that spatial downsampling benefits CE of large size of images, and gray-level downsampling accelerates histogram-dependent mapping function generation.

### *2.2 Calibration of Mapping Function*

Let the low-resolution mapping function derived from $h(k)$ be denoted by $y=m(k)$, $k=0, 1, ..., N_g-1$. Mapping function calibration aims to reconstruct a proper full dynamic mapping function $y=m^{'}(x)$, $x=0, 1, ..., 2^B-1$ from $y=m(k)$. Generally, there exist two candidate image transform schemes which are discussed detailedly as follows.

1) As Matlab function *imhisteq*, $y=m(k)$, $k=0, 1, ..., N_g-1$ is upsampled into $y=m^{'}(x)=m(\lfloor x/\Delta \rfloor)$, $x=0, 1, ..., 2^B-1$ by nearest neighboring interpolation. Then $y=m^{'}(x)$ is applied to $\mathbf{X}_s$, and the result is reversely upsampled to yield an enhanced image **Y** with the same size as **X**. Note that in some CE algorithms, such as SMIRANK, the $y=m^{'}(x)$ here can not be directly applied to **X**, since some kinds of gray levels in **X** may be missed by the spatial downsampling and excluded from histograms. Nevertheless, annoying stratification artifacts are easily incurred in **Y** due to degraded gray-level resolution, especially in the X with large smooth regions and histogram peaks. As a result, such a transform scheme is undesirable.

2) $y=m(k)$ is linearly completed and upsampled into $y=m^{'}(x)$ for covering all the gray levels of **X**, and then applied to **X** for yielding an enhanced image **Y**. The stratification artifacts can be attenuated efficiently in this scheme, which is adopted as mapping function calibration in our proposed CE acceleration framework.

Specifically, in order to reduce additional computational burden, the simple yet efficient one-dimensional linear interpolation is used to implement the upsampling and completing of $y=m(k)$. In terms of generation methods of mapping function, the completing operation may be selectively used to estimate the gray levels missed by either spatial or gray-level downsampling, as that enforced in the accelerated SMIRANK algorithm (see Section 3.1).

## 3. Case Studies on HE and SMIRANK

This section presents the case studies of our acceleration scheme on two typical CE algorithms: HE and SMIRANK. For the input $B$-bit grayscale image **X**, CE algorithms aim to yield an enhanced image **Y** with higher contrast and less distortion than **X**. As the prior works [2, 3, 6], CE of color images is realized by applying CE to luminance channel images and preserving the chrominance components in HSV color space.

### *3.1 Acceleration of HE*

The accelerated HE algorithm is proposed as follows,

*(1)* Spatially downsampling $\mathbf{X}$ to $\mathbf{X}_s$ as Eq. (1).

(*2*) Compute the downsampled gray-level histogram $h$ of $\mathbf{X}_s$ as Eq. (2).

(*3*) Obtain cumulative distribution function (CDF) $c$ from $h$.

(4) Calibrate $c$ to $c'$ with $2^B$ items by upsampling it with linear interpolation.

(*5*) Perform pixel value transform $\mathbf{Y}(i, j)=[(2^B-1)\cdot c'(\mathbf{X}(i, j))]$, where $[\cdot]$ means converting to unsigned $B$-bit integers.

Comparing with the baseline HE algorithm [1], the integrated acceleration measures are included in the steps (*1*), (*2*), (*4*). Specifically, in the step (*1*), spatial image *downsampling* decreases the number of counted pixels, which can accelerate the generation of histogram. The shorter histogram ($N_g<2^B$) yielded in the step (*2*) benefits the fast calculation of CDF. In the *calibration* phase, i.e., the step (*4*), the computationally cost-effective upsampling of CDF is used to reconstruct a continuous mapping function covering the full dynamic range $[0, 2^B-1]$ of $\mathbf{X}$.

### *3.2 Acceleration of SMIRANK*

The accelerated SMIRANK algorithm is proposed as follows,

(*1*) Compute 2D joint-spatial histogram $h_b(k)$ of the spatially downsampled image $\mathbf{X}_s$ as Eqs. (1)(2). Here, $b$=1, 2, ..., $N_b$ are the indexes of divided non-overlapped image blocks in $\mathbf{X}_s$, and $k$=0, 1, ..., $N_g$-1 are the bin indexes of blockwise gray level histograms.

(*2*) Normalize $h_b(k)$ to be $h_b(k)/((\lfloor M/s \rfloor)(\lfloor N/s \rfloor))$.

(*3*) Compute the mutual spatial information as

$$\mathbf{I}(k,l)=\sum_{b=0}^{N_b} h_b'(k,l)\log\frac{h_b'(k,l)}{h_b(k)\,h_b(l)} \tag{3}$$

where $h'_b(k, l) = \min(h_b(k), h_b(l))$, $k, l \in \mathfrak{X}=\{x_n \mid \sum_b h_b(x_n)>0,\ n=1, 2, ..., K\}$. $K$ denotes the number of non-zero columns within the matrix $\mathbf{H}$ which consists of $\mathbf{H}_{b,\,k+1}=h_b(k)$.

(*4*) Calculate $\mathbf{G}=\alpha\mathbf{S}+(1-\alpha)\mathbf{o}\mathbf{v}^T$, where $\mathbf{S}$ is created by normalizing each column of $\mathbf{I}$. $\mathbf{o}\in\mathbf{R}^{K\times 1}$ is unit vector and $\mathbf{v}\in\mathbf{R}^{K\times 1}$ is uniform vector, i.e., $\mathbf{o}^T\mathbf{v}=1$. $\alpha\in[0,1]$ is the adjustable damping factor.

($5$) The gray-level rank vector $\mathbf{r} \in \mathbf{R}^{K\times 1}$ is gained by solving $\mathbf{Gr} = \mathbf{r}$, where $\mathbf{r} = (1-\alpha)(\mathbf{E}-\alpha\mathbf{S})^{-1}\mathbf{v}$, and $\mathbf{E}$ is $K\times K$ identity matrix.

($6$) Map input gray-levels in $\mathcal{X}$ to output

$$y_k = y_{k-1} + \Delta_{k-1,k} \cdot (2^B-1) \tag{4}$$

where $k$=2, 3, ..., $K$ and $y_1$=0. $\Delta_{k-1,k} \in [0,1]$ is defined as

$$\Delta_{k-1,k} = \frac{\mathbf{r}(k-1)+\mathbf{r}(k)}{2} + \frac{\mathbf{r}(1)+\mathbf{r}(K)}{2(K-1)}. \tag{5}$$

($7$) Complete $F(x_k)=y_k$, $k$ =1, 2, ..., $K$, to $F(k)$, $k$=0, 1, ..., $N_g$-1, by filling the lost items via linear interpolation. Then $F$ is linearly upsampled to be with $2^B$ items, i.e., $F(x)$, $x$=0, 1, ..., $2^B$-1. Lastly, output $\mathbf{Y}(i,j)=[(2^B-1)\cdot F(\mathbf{X}(i,j))]$.

Comparing with SMIRANK [2], the main changes lie in the steps ($1$), ($7$). Specifically, both spatial and gray-level *downsampling* are implemented to speed up the generation of 2D joint-spatial histograms in the step ($1$). The low-dimensional blockwise histograms $h_b$ could evidently speed up the post computations of mutual information $\mathbf{I}$ and gray-level rank vector $\mathbf{r}$, which correspond to the steps ($3$) and ($5$), respectively. Such two steps constitute the main part of SMIRANK, and serve to generate an incomplete low-resolution mapping function $\{F(x_k)=y_k \mid k=1, 2, \ldots, K\}$ in the step ($6$). In the *calibration* step ($7$), $F$ is completed and upsampled by linear interpolation to recover the full dynamic mapping function $\{F(x) \mid x=1, 2, \ldots, 2^B-1\}$. Such processing refers to mapping function calibration in the acceleration framework. Lastly, the enhanced image $\mathbf{Y}$ is outputted via pixel value mapping.

## 4. Experimental Results and Discussion

### *4.1 Datasets, Algorithms and Performance Measures*

Test images are collected from four standard databases, i.e., TID2013 [17], CSIQ [18], CCID2014 [10] and RGB-NIR [19]. TID2013 image dataset includes 25 reference images and their contrast-changed versions at Levels 1~5, which respectively correspond to small contrast decreasing/increasing, larger contrast decreasing/increasing and the largest contrast decreasing. In CSIQ, 30 reference images are degraded at 5 consecutive levels, where the Levels 1 and 5 signify the smallest and largest contrast degrading, respectively. CCID2014 consists of 15 representative Kodak images [20] and their 655 contrast-distorted copies. RGB-NIR image dataset has 477 images captured in RGB and near-infrared (NIR), where RGB images are used in our tests. The size of TID2013, CSIQ, CCID2014 and RGB-NIR images are 512×384, 512×512, 768×512 and 1024×(620~768) pixels, respectively. CE algorithms are applied to the corresponding 8-bit luminance images in HSV color space.

Our proposed fast HE (FHE) and fast SMIRANK (FSMIRANK) are compared with naive HE [1], SMIRANK [2], HMF [3], AGCWD [12] and SECE [6]. Default parameter settings of primary algorithms are used. Without loss of generality, $\alpha = 0.9$ is set in both SMIRANK and FSMIRANK.

Currently, the performance assessment of image CE algorithms is still a challenge task [2, 10, 21-23]. In order to keep consistency with prior works, the state-of-art CE metrics, QRCM (quality-aware relative contrast measure) [2] and BIQME (blind image quality measure of enhanced images) [23] are used in all the tests. QRCM is a full-reference image quality assessment for measuring both the contrast change and image quality degradation between input and outputs. It yields a number within [-1, 1], where -1 and 1 denote the full level of contrast degradation and improvement, respectively. As a no-reference CE metric, BIQME captures five influencing factors: contrast, sharpness, brightness, colorfulness and naturalness. A larger BIQME score signifies better visual perceptual quality.

### *4.2 Effectiveness Evaluation of Acceleration Strategies*

Perceptual quality of enhanced images and average processing time per image are two important criteria for evaluating CE methods. In order to evaluate the visual quality change incurred by proposed accelerations, the QRCM difference between the images yielded by accelerated and naive algorithms ($\Delta QRCM = QRCM_{fast} - QRCM_{naive}$) is computed for each RGB-NIR image. CDF statistics of such $\Delta$QRCM values are plotted in Fig. 2, where (a)(c) and (b)(d) correspond to the HE and SMIRANK groups, respectively.

We also investigate the performance varying with $N_g$ and $s$. Fig. 2(a)(b) plots $\Delta$QRCM statistics varying with $N_g$ for HE and SMIRANK groups, respectively. For HE, $\Delta$QRCM values of nearly all samples are above a rather small value, i.e., -0.005. Moreover, more than 60% sample values are positive, which signifies higher QRCM gained by FHE. Overall, visual quality of the image enhanced by FHE is comparative or even better than that of HE, and is insensitive to $N_g$. Fig. 2(b) shows that QRCM values of more than 90% samples for FSMIRANK are lower than those for SMIRANK. However, there are about 99%, 90%, 70% sample values above -0.01 (a negligible degradation) for $N_g$=128, 64, 32, respectively. Such results verify that FSMIRANK is comparative or slightly worse than SMIRANK on the visual quality of outputs.

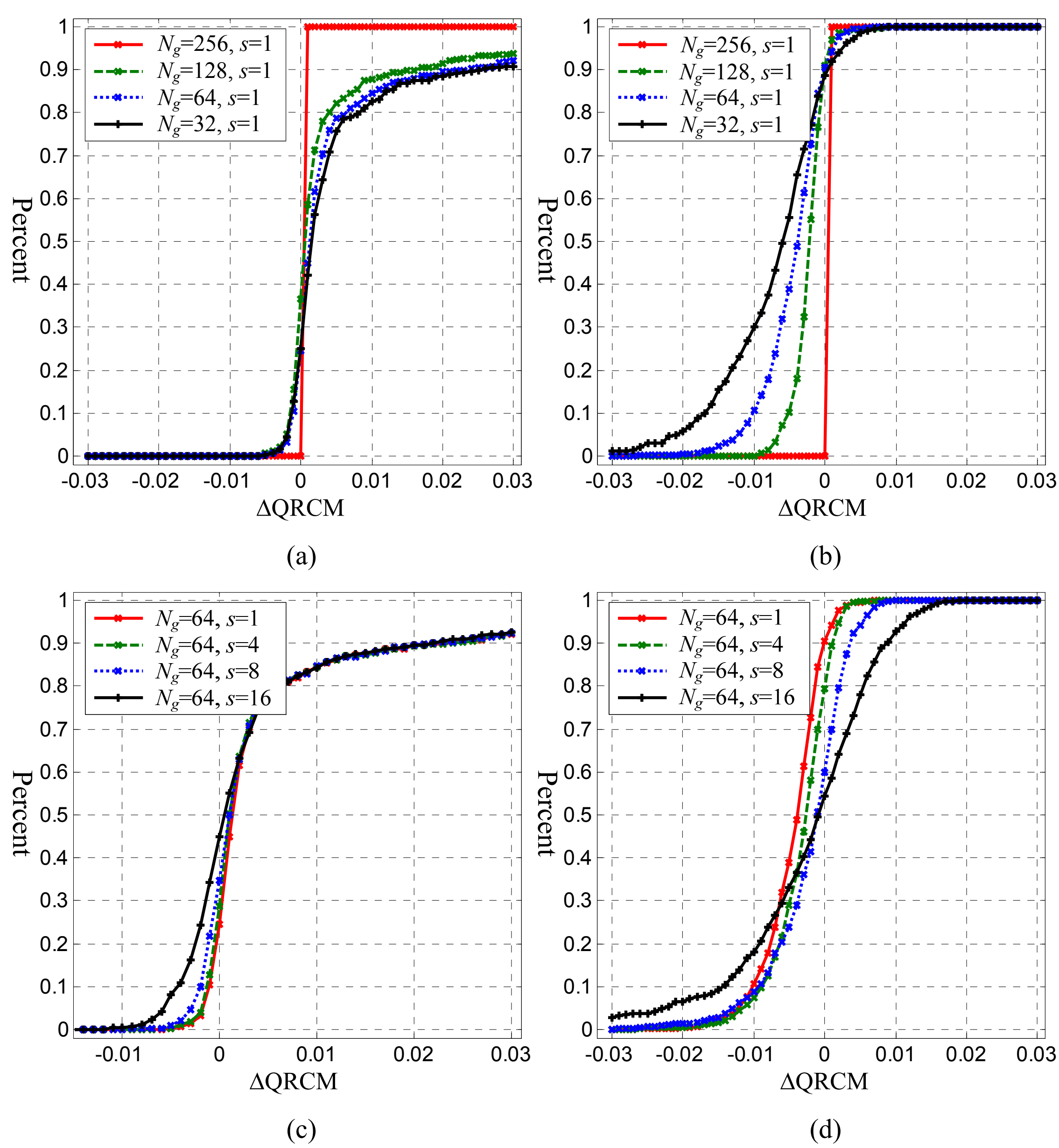

**Fig. 2.** Cumulative distribution of the difference between QRCM values yielded by (a)(c) FHE and HE; (b)(d) FSMIRANK and SMIRANK on RGB-NIR dataset under different $s$ and $N_g$ settings.

TABLE I

AVERAGE COMPUTATION TIME (MS) PER IMAGE ON RGB-NIR DATASET FOR FHE AND FSMIRANK UNDER DIFFERENT $s$ AND $N_g$ SETTINGS.

| $N_g$ | 256 | 128 | 64 | 32 | 64 | | | |
|---|---|---|---|---|---|---|---|---|
| $s$ | 1 | | | | 1 | 4 | 8 | 16 |
| **FHE** | 48.6 | 42.4 | 36.4 | 32.7 | 36.4 | 12.6 | 11.7 | 11.0 |
| **FSMIRANK** | 530.2 | 156.0 | 86.3 | 64.9 | 86.3 | 34.5 | 35.3 | 29.7 |

The computational complexity decrease incurred by accelerations is tested by running CE algorithms on a computer with Intel Core i5-5200U CPU @ 2.2 GHz and 8G RAM under MATLAB R2013a. As shown in Table I, the average processing time for one image decreases monotonously with $N_g$. Without loss of generality, $N_g$=64 is selected for limiting the degradation and computational complexity.

Results on the visual quality varying with $s$ are shown in Fig. 2(c)(d). Overall, the performances at $s$=1, 4, 8 are comparative, and far better than that at $s$=16. Table I shows that the computing speed also decreases monotonously with $s$. In terms of the trade-off between computation cost and visual quality, $s$=8 is selected. In conclusion, the effectiveness of our proposed CE acceleration strategies can be verified by such baseline evaluation results.

### *4.3 Comparing with Other CE Methods*

Objective and subjective performance assessment of the accelerated CE methods is also enforced by comparing with other CE algorithms on extensive datasets. Table II shows the average QRCM values for different CE algorithms on four databases. It shows that FHE is comparative or slightly better than HE, where the increment falls within [-0.003, 0.044]. Among all methods, SMIRANK ranks first and is nearly followed by FSMIRANK, which owns comparative or slightly lower QRCM values. The decrement falls within [-0.01, 0]

TABLE II

AVERAGE QRCM (x$10^{-2}$) FOR DIFFERENT CE ALGORITHMS ON EACH DATASET. HERE, $s$=8, $N_g$=64. LEVEL 0 MEANS UNALTERED IMAGES. THE LARGEST TWO PER ROW ARE LINED.

| **Algorithm** | | **HMF** | **AGCWD** | **SECE** | **HE** | **FHE** | **SMIRANK** | **FSMIRANK** |
|---|---|---|---|---|---|---|---|---|
| **TID2013** | Level 0 | 10.1 | 9.1 | 7.5 | 10.7 | 12.0 | 13.5 | 13.5 |
| | Level 1 | 11.5 | 12.3 | 11.8 | 15.8 | 15.5 | 17.5 | 17.3 |
| | Level 2 | 7.7 | 5.6 | 4.0 | 4.6 | 7.2 | 9.3 | 9.3 |
| | Level 3 | 14.1 | 17.8 | 18.8 | 22.1 | 22.0 | 23.9 | 23.7 |
| | Level 4 | 5.5 | 2.7 | 1.8 | -1.9 | 2.5 | 5.9 | 5.9 |
| | Level 5 | 17.1 | 28.8 | 31.2 | 33.0 | 32.8 | 34.9 | 34.9 |
| **CSIQ** | Level 0 | 7.6 | 6.3 | 3.1 | 3.3 | 3.4 | 7.9 | 7.7 |
| | Level 1 | 10.5 | 10.5 | 9.3 | 9.3 | 9.1 | 13.7 | 13.2 |
| | Level 2 | 14.2 | 17.0 | 18.1 | 17.8 | 17.6 | 21.9 | 21.3 |
| | Level 3 | 17.5 | 30.0 | 33.0 | 31.5 | 31.5 | 35.0 | 34.1 |
| | Level 4 | 17.8 | 35.0 | 37.7 | 36.1 | 36.1 | 38.9 | 37.9 |
| | Level 5 | 17.8 | 35.0 | 37.7 | 36.1 | 36.1 | 38.9 | 37.9 |
| **CCID2014** | | 13.0 | 9.8 | 11.9 | 11.8 | 13.9 | 17.3 | 17.4 |
| **RGB-NIR** | | 12.9 | 9.4 | 7.8 | 11.0 | 11.7 | 14.9 | 14.7 |

TABLE III

AVERAGE BIQME ($\times 10^{-1}$) FOR DIFFERENT CE ALGORITHMS ON EACH DATASET. HERE, $s$=8, $N_g$=64. UNEN MEANS UNENHANCED IMAGES. THE LARGEST TWO PER ROW ARE LINED.

| Algorithm | | UNEN | HMF | AGCWD | SECE | HE | FHE | SMIRANK | FSMIRANK |
|---|---|---|---|---|---|---|---|---|---|
| **TID2013** | Level 0 | 59.0 | 63.0 | 60.9 | 61.6 | 64.1 | 64.6 | 63.5 | 63.7 |
| | Level 1 | 56.3 | 61.5 | 59.1 | 61.1 | 64.4 | 64.1 | 63.2 | 63.3 |
| | Level 2 | 62.3 | 64.8 | 63.8 | 63.4 | 64.1 | 65.1 | 64.7 | 64.9 |
| | Level 3 | 51.5 | 58.5 | 56.3 | 60.4 | 64.2 | 64.0 | 62.8 | 62.9 |
| | Level 4 | 63.7 | 65.2 | 65.0 | 64.1 | 63.1 | 64.8 | 64.8 | 65.0 |
| | Level 5 | 41.9 | 50.8 | 51.0 | 59.2 | 63.9 | 63.5 | 62.3 | 62.2 |
| **CSIQ** | Level 0 | 60.1 | 62.8 | 62.4 | 61.4 | 65.2 | 65.3 | 63.4 | 62.6 |
| | Level 1 | 55.6 | 60.5 | 58.9 | 59.4 | 62.8 | 62.8 | 61.7 | 61.7 |
| | Level 2 | 51.2 | 57.8 | 56.7 | 58.6 | 62.1 | 62.2 | 61.4 | 61.4 |
| | Level 3 | 42.2 | 49.6 | 52.2 | 57.8 | 61.3 | 61.3 | 61.0 | 60.8 |
| | Level 4 | 38.0 | 45.0 | 49.6 | 57.2 | 60.4 | 59.9 | 60.5 | 60.1 |
| | Level 5 | 38.0 | 45.0 | 49.6 | 57.2 | 60.4 | 59.9 | 60.5 | 60.1 |
| **CCID2014** | | 50.1 | 57.3 | 55.2 | 57.9 | 63.2 | 63.9 | 61.1 | 60.8 |
| **RGB-NIR** | | 54.6 | 60.2 | 56.8 | 57.3 | 62.0 | 62.3 | 60.4 | 60.6 |

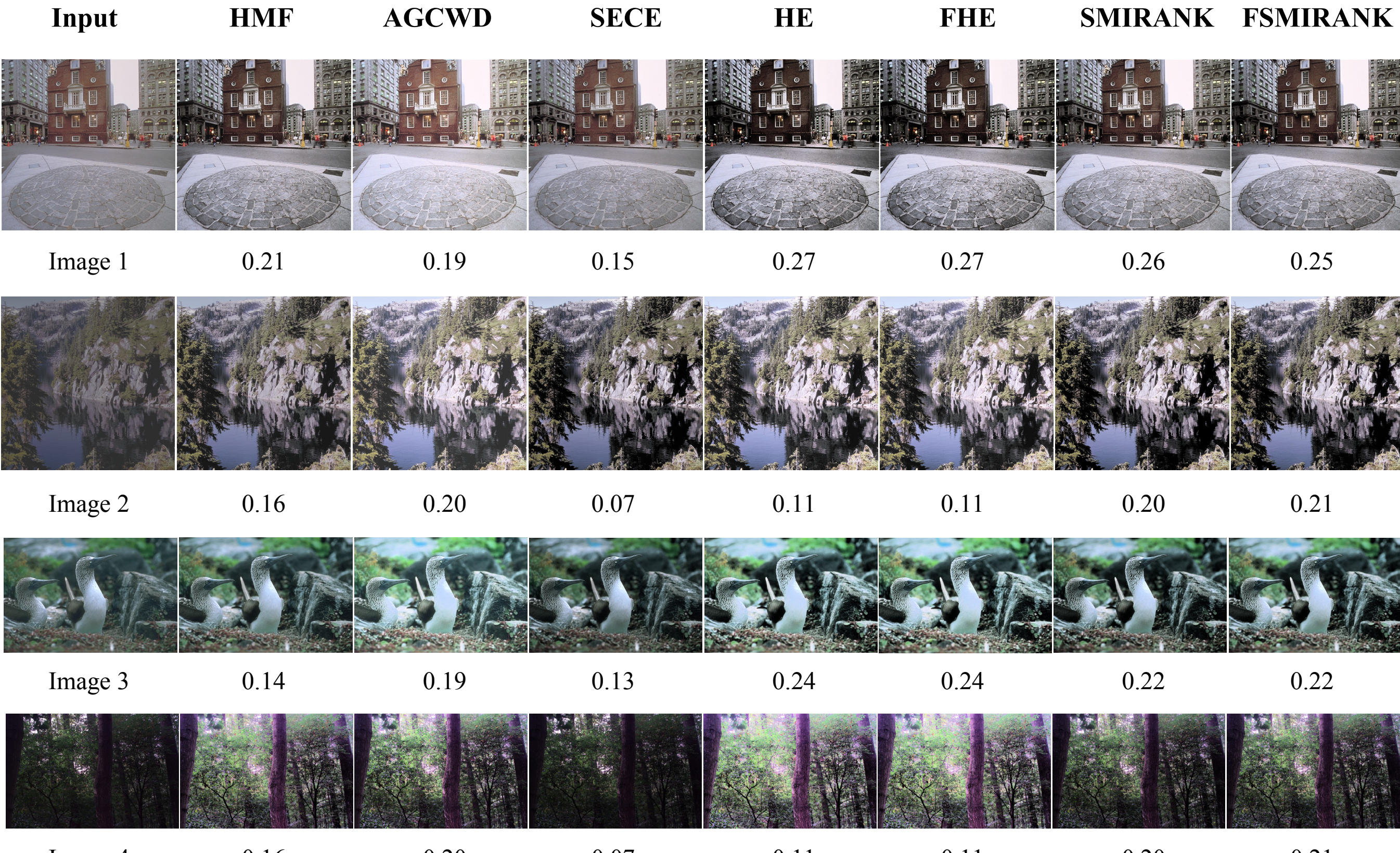

**Fig. 3.** Results for different CE methods on four example images. The corresponding QRCM value of each image is shown below.

with the average of -0.004, which is rather smaller than the margins above the other CE methods including HMF, AGCWD and SECE. Table III lists the corresponding test results achieved by the BIQME metric. Although the measurements are prone to HE, the consistent conclusion as that of QRCM could also be obtained. All such quantitative and objective assessment results verify that the accelerated methods can preserve the visual enhancement quality of naive CE methods, and generally behave better than HMF, AGCWD and SECE.

Fig. 3 shows the qualitative visual quality assessment enforced on four example images. The images 1-2, 3, 4 are from CSIQ, BSD500 [24] and RGB-NIR databases, respectively. From both visual observation and QRCM measurements, we can see that perceptual quality of the enhanced images yielded by accelerated methods is preserved successfully. Except for those inherently brought by naive CE methods, no additional unnatural artifacts would be incurred by the corresponding accelerated methods.

### *4.4 Complexity Comparison*

Time complexity of different CE algorithms is also tested. The average processing time for each image of the four different datasets is shown in Table IV, which indicates that FHE outperforms other methods remarkably. Although HE is famous for its fast processing speed, it is still speeded up by about 3.9 times by our proposed acceleration scheme. Moreover, SMIRANK is impressively speeded up by about 13.5 times by FSMIRAMK. Such evident improvement should attribute to both the spatial and gray-level downsampling. FSMRANK runs much faster than SECE and AGCWD. Comparing with HMF, FSMIRANK behaves comparatively on TID2013 and CSIQ, but better on CCID2014 and RGB-NIR. Such results should attribute to the benefit of our proposed methods on enhancing the relatively large size of images, such as those from RGB-NIR and CCID 2014.

We also analyze the theoretical time complexity of CE algorithms in enhancing a $M \times N$ $B$-bit grayscale image. The analysis results are shown in Table V. For HE, computing the

TABLE IV

AVERAGE COMPUTATION TIME (MS) PER IMAGE FOR DIFFERENT CE ALGORITHMS ON EACH DATASET. HERE, $S$=8, $N_g$=64. THE FASTEST TWO PER ROW ARE LINED.

| Algorithm | HMF | AGCWD | SECE | HE | FHE | SMIRANK | FSMIRANK |
|---|---|---|---|---|---|---|---|
| **TID2013** | 18.0 | 37.2 | 54.3 | 14.7 | 4.0 | 314.0 | 22.6 |
| **CSIQ** | 23.9 | 50.7 | 60.8 | 19.6 | 5.3 | 283.0 | 21.0 |
| **CCID2014** | 34.8 | 68.8 | 75.7 | 29.1 | 6.9 | 307.8 | 25.5 |
| **RGB-NIR** | 56.9 | 118.6 | 116.4 | 47.4 | 11.7 | 519.4 | 35.3 |

TABLE V

THEORETICAL TIME COMPLEXITY OF NAIVE AND ACCELERATED CE ALGORITHMS.

| Algorithm | Time Complexity |
|---|---|
| **HE** | $O(2MN+2^B)$ |
| **FHE** | $O((1/s^2+1)MN+N_g)$ |
| **SMIRANK** | $O(2MN+K2^{2B}+K^3)$ |
| **FSMIRANK** | $O((1/s^2+1)MN+KN_g{}^2+(KN_g/2^B)^3)$ |

histogram requires time $O(MN)$. Generating the mapping function requires time $O(2^B)$, and finally applying pixel-wise transform to yield the enhanced image requires time $O(MN)$. As a result, the total time complexity of HE is $O(2MN+2^B)$ [3]. For FHE, such corresponding three items require times $O(MN/s^2)$, $O(N_g)$ and $O(MN)$, respectively. The mapping function calibration costs time $O(1)$. So the total time complexity of FHE is $O((1/s^2+1)MN+N_g)$.

For SMIRANK, the whole computational cost mainly comes from the computations of blockwise histograms, mutual information matrix, the rank vector and pixel value mapping. Such four items requires times $O(MN)$, $O(K2^{2B})$, $O(K^3)$ and $O(MN)$, respectively. Here, $K$ denotes the number of gray levels existing in the primary input image. Therefore, the total time complexity for SMIRANK is $O(2MN+K2^{2B}+K^3)$. Correspondingly, the total time complexity of FSMIRANK is $O((1/s^2+1)MN+KN_g{}^2+(KN_g/2^B)^3)$, where the mutual information matrix and rank vector are calculated in terms of downsampled gray levels.

We also noted that $\alpha$ can be set automatically based on the gradient magnitude map of input images [2]. We also conduct related experiments and find that such automatic setting may slightly improve the visual quality at a cost of increasing a little computational time. Nevertheless, comparing with the prior art, the computational performance superiority of FSMIRAMK gained by the integrated acceleration strategies is still rather evident.

## 5. Conclusions

In this paper, a fundamental framework is proposed to accelerate general histogram-based image CE algorithms. Both spatial downsampling and histogram simplifying mechanisms are investigated deeply and adapted to significantly decrease the computational complexity of prior CE techniques. Mapping function calibration is novelly proposed to reconstruct the transform on the gray levels missed by downsampling. The case studies on two typical CE algorithms, i.e., HE and SMIRANK, are presented detailedly. Effectiveness of our proposed CE acceleration scheme has been validated by the extensive experimental results on four standard databases. In conclusion, our proposed CE acceleration framework can be friendly used to remarkably improve the computational efficiency of histogram-based CE algorithms, while perceptual quality of enhanced images can still be preserved.